\documentclass[10pt,preprintnumbers,showpacs,amsmath,amssymb,floatfix,twocolumn,prb]{revtex4} 

\usepackage{graphicx}
\usepackage{dcolumn}
\usepackage{bm}
\usepackage{longtable}
\usepackage{graphics}
\usepackage{xspace}
\usepackage{epstopdf}
\usepackage{epsfig}
\newcommand{\ud}{$U_d$ }
\newcommand{\udn}{$U_d$}
\newcommand{\umv}{$U_m^{(v)}$ }

\newcommand{\vmvn}{$V_m^{(v)}$}
\newcommand{\udv}{$U_d^{(v)}$ }
\newcommand{\udvn}{$U_d^{(v)}$}
\newcommand{\xid}{$\xi_d$ }
\newcommand{\xidn}{$\xi_d$}
\newcommand{\bX}{$\beta$-(ET)$_2X$ }

\newcommand{\bI}{$\beta$-(ET)$_2$I$_3$ }
\newcommand{\bIn}{$\beta$-(ET)$_2$I$_3$}
\newcommand{\Br}{$\kappa$-(ET)$_2$Cu[N(CN)$_2$]Br }

\newcommand{\kXn}{$\kappa$-(ET)$_2X$}

\newcommand {\etal}{{\it et al}. }
\newcommand {\etalc}{{\it et al}., }
\newcommand {\etaln}{{\it et al}.}

\begin{document}
\title{Effective Coulomb interactions within BEDT-TTF dimers}
\author{Edan Scriven}
\email{edan@physics.uq.edu.au}
\author{B. J. Powell}
\affiliation{Centre for Organic Photonics and Electronics, School of Mathematics and Physics, University of Queensland 4072, Australia}

\begin{abstract}
We calculate the effective Coulomb interactions between  holes in dimers of the organic molecule BEDT-TTF \textit{in vacuo}. We use density functional theory (DFT) to parameterise Hubbard models for $\beta$ and $\kappa$ phase organic charge transfer salts. We focus  on the intra-dimer Coulomb repulsion, \udvn, and the inter-monomer Coulomb repulsion, $V_m^{(v)}$. We find that \udv $= 3.22 \pm 0.09$ eV and $V_m^{(v)} = 2.71 \pm 0.10$ eV for 23 experimental geometries taken from a range of materials in both the $\beta$ and $\kappa$ polymorphs. The quoted error is one standard deviation over the set of conformations studied. We conclude that \udv and $V_m^{(v)}$ are essentially the same for an isolated dimer with the geometries present in all of the compounds studied. We highlight the disagreement between our parameterisation of the Hubbard model and previous results from both DFT and H\"{u}ckel methods and show that this is caused by the failure of an assumption made in previous  calculations (that $U_m^{(v)}\gg V_m^{(v)}$, where $U_m^{(v)}$ is the effective intra-monomer Coulomb repulsion). We discuss the implications of our calculations for theories of the BEDT-TTF salts based on the Hubbard model on the 2D anisotropic triangular lattice and explore the role of conformational disorder in these materials.
\end{abstract}

\maketitle

\section{Introduction}

Layered organic charge transfer salts of the form (ET)$_2X$, where ET is bis(ethylenedithio)tetrathiafulvalene or BEDT-TTF and $X$ is a monovalent anion, exhibit a variety of unusual phenomena due to the strong electronic correlations present in these materials.\cite{Powell:strcor_review} These phenomena include unconventional superconductivity\cite{Powell:strcor_review} with a small superfluid stiffness,\cite{PowellPratt} a Mott insulator,\cite{Kanoda} a spin liquid,\cite{ShimuzuLee} strongly correlated\cite{Jacko,JaimeLimellette} and unconventional\cite{JaimeLimellette,Nam} metallic states, and a pseudogap.\cite{Eddy} Experimentally, one can tune between these phases by varying the temperature and pressure (including both hydrostatic and `chemical' pressure, i.e., varying the anion, $X$).\cite{Powell:strcor_review}

DFT, as implemented with current approximate exchange-correlation functionals, does not capture several important aspects of  the physics of strongly-correlated electronic systems.\cite{YangScience} For example, DFT band structure calculations of ET crystals,\cite{Lee:DFT_band,Kandpal:DFT_band,Nakamura:DFT_band}  produce a half-filled valence band and hence a metallic state. However, these calculations do not recover the Mott insulating state or the other strongly correlated effects that are observed experimentally. Therefore, efforts have focused on the application of effective low-energy Hamiltionians, such as Hubbard models.\cite{Powell:strcor_review} However, in molecular crystals, the effective parameters for such low energy Hamiltonians may be calculated from studying the properties of single molecules or small molecular clusters, which may be accurately described by DFT.\cite{Martin,Antropov,Quong,Brocks,Laura,Scriven:monomers,BJPchapter}

ET salts occur in a variety of crystal packing structures. In the $\beta$ and $\kappa$ polymorphs the ET molecules form dimers. Intradimer dynamics are often integrated out of effective low energy models of \bX and \kXn. In these charge transfer salts, each dimer donates one electron to the anion layers to form a half-filled system. Both H\"uckel \cite{Canadell} and DFT \cite{Lee:DFT_band,Kandpal:DFT_band,Nakamura:DFT_band} calculations have found that the dimers form an anisotropic triangular lattice in which each lattice site is a dimer. However, there is a strong effective Coulomb repulsion, \udn, between two electrons on the same dimer, which must be included in the effective low energy description.\cite{Powell:strcor_review,KinoFukuyama}  The electronic interactions within an ET dimer are stronger than those between an ET molecule and its next-nearest neighbours on the crystal lattice. Therefore, these materials have been widely studied on the basis of Hubbard models.\cite{Powell:strcor_review} In order to explain the observed physics these theories assume that both chemical and hydrostatic pressure reduce $U_d/W$, where $W$ is the bandwidth. Therefore, an important task for the field is to understand how this ratio varies with chemical and hydrostatic pressure.

Previously, the on-site Coulomb repulsion term in the Hubbard model, \udn, has been estimated from both H\"{u}ckel\cite{H-lit:Fortunelli,Emge,H-lit:Mori,H-lit:Schlueter,H-lit:Simonov,H-lit:Campos,kNCS2:Rahal,H-lit:Komatsu,kI3:Kobayashi} and density functional\cite{Kandpal:DFT_band,Nakamura:DFT_band}  calculations under the assumptions that the intra-monomer Coulomb repulsion $U_m \rightarrow \infty$, and the inter-monomer Coulomb repulsion $V_m \rightarrow 0$. We will show below that this assumption is incorrect and leads to a systematic underestimate of \udn.

Disorder plays a number of important roles in organic superconductors.\cite{Powell:disorder} Increasing the degree of disorder leads to a suppression of the superconducting critical temperature, $T_c$, which is correlated with a rise in the residual resistivity.\cite{Powell:disorder} Further disorder can cause a violation of Matthiessen's rule via impurity assisted tunnelling in the interlayer direction.\cite{Analytis} 

In \Br the degree of disorder can be increased by increasing the rate at which the sample is cooled,\cite{Taylor,Taniguchi,Stalcup,Su,Su2} which leads to a suppression of $T_c$ by $\sim1$ K. Further, increasing the cooling rate can drive the system towards the insulating side of the metal-insulator transition.\cite{Taniguchi} Two hypothesis have been proposed for the source of this disorder:  terminal ethylene group disorder\cite{Taniguchi,Stalcup,Su,Su2} and disorder in the anion layer.\cite{Wolter} Therefore it is important to estimate the scattering rate caused by terminal ethylene disorder in this material.

Even more dramatic effects are observed in \bIn. Variations of the pressure as the sample is cooled can change the ambient pressure $T_c$ from 1 K (for samples cooled at ambient pressure; known as the $\beta_L$  phase) to 7 K (for samples cooled at $P\gtrsim1$ kbar once the pressure is released; known as the $\beta_H$ phase).\cite{Ginodman} In this material clear differences in the terminal ethylene groups in the $\beta_H$ and $\beta_L$ phases are observed via x-ray  scattering.\cite{Ravy} Thus it has been argued that the terminal ethylene disorder is responsible for the differences in the critical temperatures between the $\beta_H$ and $\beta_L$ phases.\cite{Powell-JPIV}

Therefore, we also present calculations of the effective site energy for holes,  $\xi_d$, for the $\beta$- and $\kappa$- phase salts. This allows us to study the effects of impurity scattering caused by conformational disorder in the terminal ethylene groups of the ET molecule.

In this paper we present DFT calculations for ET dimers in vacuum. In Sec. II we describe the computational method by which we calculate these energies. In Sec. III we discuss the problem for the isolated dimer and review the parameterisation of the two-site extended Hubbard model from the total energies of the relevant (ET)$_2$ charge states. In Sec. IV we report and discuss the resulting values of \udvn, $V_m^{(v)}$ and $\xi_d$. In Sec. V we draw our conclusions.

\section{Computational methods}

We used DFT to calculate the total energies of ET dimers in various conformations and charge states. We used the SIESTA\cite{SIESTA} implementation of DFT, with the PBE exchange-correlation functional,\cite{PBE} a triple-$\zeta$ plus single polarisation (TZP) basis set (except where we explicitly specify otherwise) and basis functions consisting of Sankey type numerical atomic orbitals.\cite{Sankey} The orbital functions were confined to a radius $r_c$ from their centres, which slightly increases the energy of the orbital. The specified maximum allowed increase in energy due to this cutoff was 2 mRy. The convergence of the integration mesh was determined by specifying an effective plane-wave cutoff energy of 250 Ry. The initial spin moments on each atom were arranged antiferromagnetically wherever possible. We used pseudopotentials constructed according to the improved Troullier-Martins (TM2) method.\cite{TM}

Nuclear positions for C and S atoms were obtained from x-ray crystallography.\cite{bI3:Leung,bAuI2:Wang,bIBr2:Williams,kI3:Kobayashi,kNCS2:Rahal,kNCN2X:Geiser,kCl3kbar:Schultz,kCN3:Geiser} H atoms, which are not observed in x-ray scattering experiments, were relaxed by the conjugate-gradient method. Total DFT energy differences between the relevant charge states $\left[E(1) - E(0)\right]$ and $\left[E(2) - E(0)\right]$ were equated with the corresponding analytical expressions in Eq. (\ref{eqn:evals}) to determine the Hubbard parameters. We focus on these `experimental' geometries rather than performing a full relaxation for a number of reasons. Firstly, there are small differences in the reported geometries for different ET salts, and one would like to understand the effect of these. Secondly, the experiments effectively `integrate over' all of the relevant charge states and therefore provide an `average' conformation. Thirdly, the experiments naturally include the effects on the molecular conformation due to the crystalline environment, which are absent from {\it in vacuo} calculations.

\section{The two site extended Hubbard model}

In calculations of the effective Coulomb interaction (the Hubbard $U$) in molecular solids it is important to recognise that their are two contributions.\cite{Martin,Antropov,Quong,Brocks,Laura,Scriven:monomers,BJPchapter} That is, the effective Coulomb interaction on a ET dimer may be written as
\begin{equation}
U_d = U_d^{(v)} - U_d^{(p)}
\end{equation}
where \udv is the value of \ud for the dimer cluster in vacuum, and $U_d^{(p)}$ is the reduction in \ud from the polarisable crystalline environment. Calculating $U_d^{(p)}$ for ET salts is a highly non-trivial problem due to the large size of the ET molecule relative to the intermolecular spacing. Below we present results of DFT calculations for \udv of dimers in the conformations found in a wide range of $\kappa$ and $\beta$ phase ET salts. Similar results hold for $U_m$, the effective Coulomb repulsion between two holes on the same monomer and $V_m$ the effective Coulomb interaction between two holes on neighbouring monomers. Below we will primarily discuss the vacuum contributions to these terms, \umv and \vmvn.

The effective vacuum intradimer Coulomb energy, \udvn, is given by (see, e.g., Ref. \onlinecite{Scriven:monomers}) 
\begin{equation}
U_d^{(v)} = E(0) + E(2) - 2 E(1), \label{eqn:Ud} 
\end{equation}
where $E(q)$ is the ground state energy of the dimer in vacuum containing $q$ holes, i.e., with charge $+q$. Similarly, the effective site energy for holes is given by 
\begin{equation}
\xi_d = E(0) - E(1). \label{eqn:xid} 
\end{equation}
Below we calculate $E(q)$ via density functional methods.

It is also interesting to consider intradimer dynamics, which can be described via a two site extended Hubbard model,\cite{Powell:strcor_review}
\begin{eqnarray}
\hat{\cal H} &=& \sum_{i\sigma} \xi_{mi} \hat{n}_{i\sigma} - t \sum_{\sigma} \left(\hat{h}^{\dagger}_{1\sigma} \hat{h}_{2\sigma} + h.c.\right) \nonumber \\
&+& \sum_{i} U_{mi} \hat{n}_{i\uparrow} \hat{n}_{i\downarrow} + V_m  \hat{n}_{1} \hat{n}_{2} \label{eqn:Hamiltonian}
\end{eqnarray}
where $\hat{h}^{(\dagger)}_{i\sigma}$ annihilates (creates) a hole on site (monomer) $i$ in spin state $\sigma$, $\xi_{mi}$ is the site energy for holes on site $i$, $\hat{n}_{i\sigma}$ is the number operator for spin $\sigma$ holes on site $i$,  $\hat n_i=\sum_\sigma \hat n_{i\sigma}$,  $t$ is the intradimer hopping integral, $U_{mi}$ is the effective on-site (monomer) Coulomb repulsion, and $V_m$ is the intersite Coulomb repulsion.

\begin{figure}
\epsfig{figure=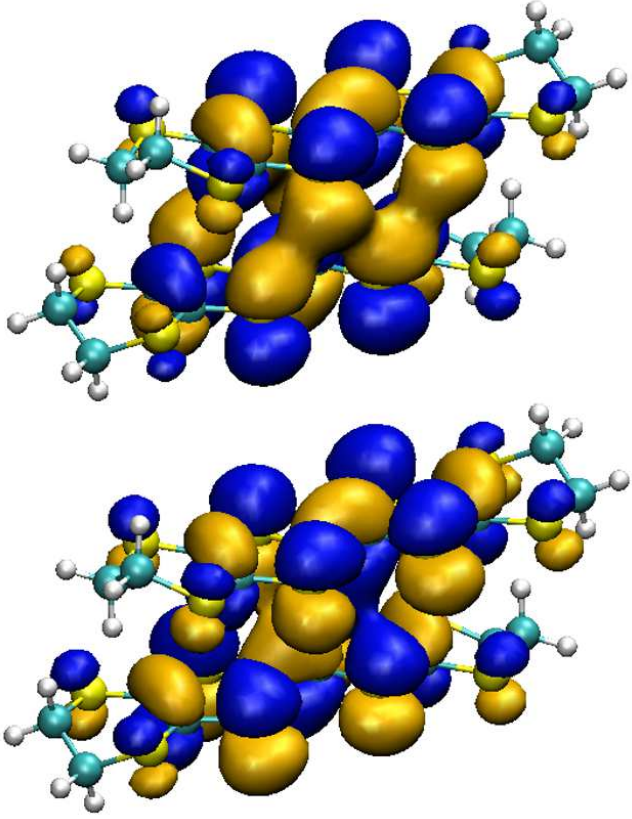, width=8.6cm}
\caption{The HOMO of (ET)$_2^{2+}$ (top) and charge neutral (ET)$_2$ (bottom), with nuclear positions from the crystal $\beta$-(ET)$_2$I$_3$. The HOMO of (ET)$_2^{2+}$ is the dimer bonding orbital and the HOMO of (ET)$_2$ is the antibonding orbital of the two ET HOMOs (cf. Fig. \ref{fig:monomer_HOMO}). The essential difference between the two lies in the relative phase of the orbital function on each molecule. The bonding orbital connects the ET molecules at the S$\cdots$S contacts  (cf. Fig. \ref{fig:2d_conformations}). In the antibonding orbital, there are nodes between the S$\cdots$S contacts. }\label{fig:b_HOMO}
\end{figure} 

\begin{figure}
\epsfig{figure=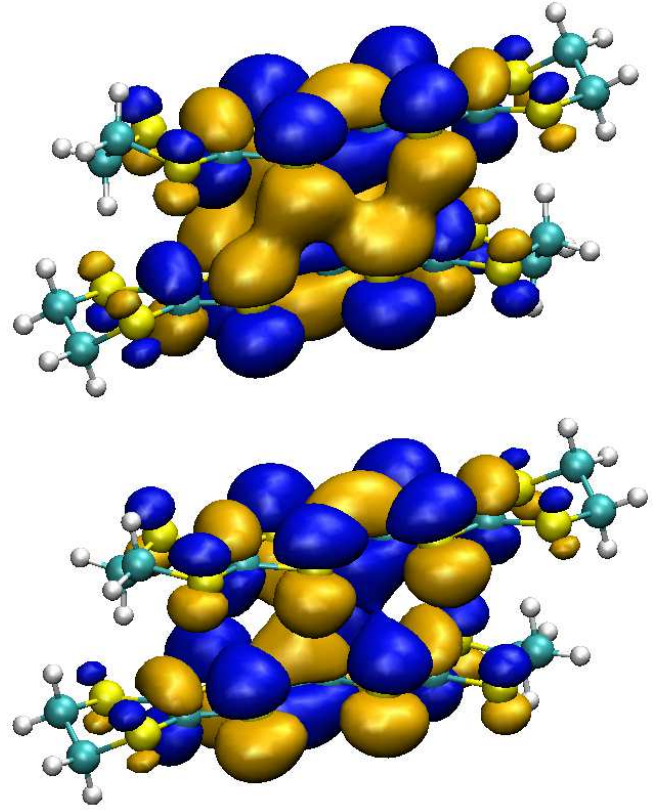, width=8.6cm}
\caption{The HOMO of (ET)$_2^{2+}$ (top) and charge neutral (ET)$_2$ (bottom), with nuclear positions from the crystal $\kappa$-(ET)$_2$Cu$_2$(CN)$_3$. The similarity of the nuclear structures and orbitals between this conformation and the $\beta$ conformation in Fig. \ref{fig:b_HOMO} highlight the dimer as a common structural unit within two different packing motifs.}\label{fig:k_HOMO}
\end{figure} 

\begin{figure}
\epsfig{figure=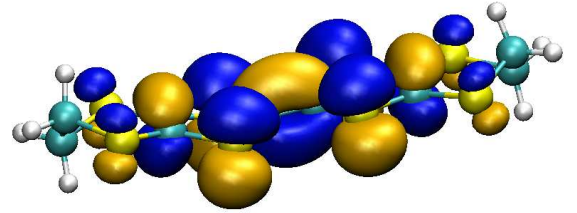, width=7.5cm}
\caption{The HOMO of a charge neutral ET monomer, with nuclear positions from the crystal $\beta$-(ET)$_2$I$_3$. This is the orbital from each molecule that contributes to the HOMO of the (ET)$_2$ and (ET)$_2^{2+}$ dimers.}\label{fig:monomer_HOMO}
\end{figure}

The lowest energy eigenvalues of Hamiltonian (\ref{eqn:Hamiltonian}) for each charge state are
\begin{subequations}\label{eqn:evals}
\begin{eqnarray}
E(0) &=& 0, \\
E(1) &=& \overline{\xi}_m - \frac{1}{2}\sqrt{4t^2 + (\Delta\xi_m)^2}, 
\end{eqnarray}
and 
\begin{eqnarray}
E(2) &=& 2\overline{\xi}_m + \frac{1}{3}\left(2\overline{U}_m + V_m - 2 A \cos \theta\right)  
\end{eqnarray}
\end{subequations}
where
$\overline{\xi}_m = \frac{1}{2}(\xi_{m1}+\xi_{m2})$, 
$A =12t^2+(\Delta U_m)^2+(U_{m1}-V_m)(U_{m2}-V_m)+3(\Delta\xi_{m})^2$,
$\cos3\theta=(\overline{U}_m-2V_m)[18t^2-(2U_{m1}-U_{m2}-V_m) (U_{m1}-2U_{m2}+V_m)-9(\Delta\xi_m)^2]/{2 A^3}$.
$\Delta\xi_m = \xi_{m1}-\xi_{m2}$, 
$\overline{U}_m = \frac{1}{2}(U_{m1}+U_{m2})$, and
$\Delta U_m = U_{m1}-U_{m2}$.

We have previously calculated $\xi_m$ and \umv from the one site Hubbard model for an ET monomer for the experimental observed conformations in all the materials discussed below,\cite{Scriven:monomers} therefore one may solve Eqs. (\ref{eqn:evals}) for $t$ and $V_m$ taking $\xi_m$ and \umv from the monomer calculations. The case of two holes with different site energies and on-site Coulomb repulsion may be solved by a general method for diagonalising cubic matrix eigensystems.\cite{Cocolicchio:cubic_eigensystems}
In cases where the two molecules in a dimer have the same geometry (e.g., by symmetry), $\xi_{m1}=\xi_{m2}=\xi_m$ and $U_{m1}=U_{m2}=U_m$ and the eigenvalues simplify to
\begin{subequations} \label{eqn:simpleEvals}
\begin{eqnarray}
E(1) &=& \xi_m - t \\
E(2) &=& 2\xi_m + \frac{1}{2}\left(U_m + V_m - \sqrt{16t^2 + (U_m - V_m)^2}\right) \nonumber \\
\end{eqnarray}
\end{subequations}
in which case the solution is straightforward.

In the limit $U_m=V_m=0$ the two site Hubbard model has two solutions: the bonding state $|\phi_{b\sigma}\rangle=|\phi_{1\sigma}\rangle+|\phi_{2\sigma}\rangle$ and the antibonding state $|\phi_{b\sigma}\rangle=|\phi_{1\sigma}\rangle-|\phi_{2\sigma}\rangle$, where $|\phi_{i\sigma}\rangle=\hat h_{i\sigma}|0\rangle$ is a single electron state centred on the $i^\textrm{th}$ monomer and $|0\rangle$ is the (particle) vacuum state.

In Figs. \ref{fig:b_HOMO} and \ref{fig:k_HOMO} we plot the HOMOs of ET dimer for the conformations found in $\beta$-(ET)$_2$I$_3$ and $\kappa$-(ET)$_2$Cu$_2$(CN)$_3$ respectively, in both the charge neutral and the 2+ states. It can be seen that these dimer orbitals are the antibonding and bonding hybrids of the ET monomer HOMO (shown in Fig. \ref{fig:monomer_HOMO}), respectively. The most important difference between the orbital geometries lies in the S$\cdots$S intermolecular contacts, which contain nodes in the antibonding orbital, but are connected in the bonding orbital.
Thus the DFT picture of the (ET)$_2$ system is remarkably similar to the molecular orbital description of a diatomic molecule,\cite{BJPchapter} but with the `covalent bond' between the two monomers rather than between two atoms.

\section{Calculation of the Hubbard model parameters}

\subsection{Basis set convergence}

\begin{figure}
\epsfig{figure=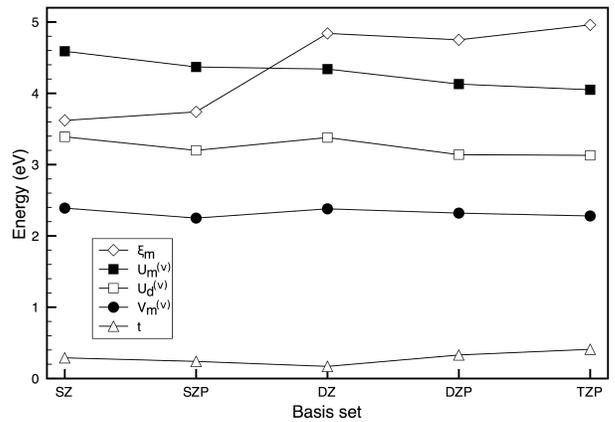, width=8.6cm}
\caption{Variation of Hubbard parameters found from DFT calculations with basis set, which improve from left to right. The test conformation is taken from the crystal $\kappa$-(ET)$_2$Cu(NCS)$_2$.\cite{kNCS2:Rahal} All of the quantities except $t$ are well-converged at TZP, the basis set chosen for all subsequent calculations. Indeed, $\xi_m$ is the only other quantity that changes significantly across the range of basis sets. However, $t$ is a relatively small quantity, on the order of its own variations with respect to basis set size. Hence we conclude that solving the Hamiltonian (\ref{eqn:Hamiltonian}) is not an accurate method for finding $t$.}\label{figure:DFT_convergence}
\end{figure} 

We tested the basis set convergence of the DFT calculations using the conformation observed in $\kappa$-(ET)$_2$Cu(NCS)$_2$ as the test case, with single-$\zeta$ (SZ), single-$\zeta$ plus polarisation (SZP), double-$\zeta$ (DZ), double-$\zeta$ plus polarisation (DZP) and TZP basis sets. We also calculated the monomer parameters, $U_m$ and $\xi_m$, in each basis, using the method we previously applied to the ET monomers.\cite{Scriven:monomers} The Hubbard model parameters in each basis set are reported in Fig. \ref{figure:DFT_convergence}. The values of all parameters are well-converged in the TZP basis, except $t$. $t$ is an order of magnitude smaller than the other parameters, and on the order of both the variation of the other parameters among the basis sets tested and the uncertainty associated with the calculation method. This suggests that extracting $t$ from band structure calculations\cite{Kandpal:DFT_band,Nakamura:DFT_band} is a more accurate and reliable method of estimating the hopping integrals in these systems.

\begin{figure}
\epsfig{figure=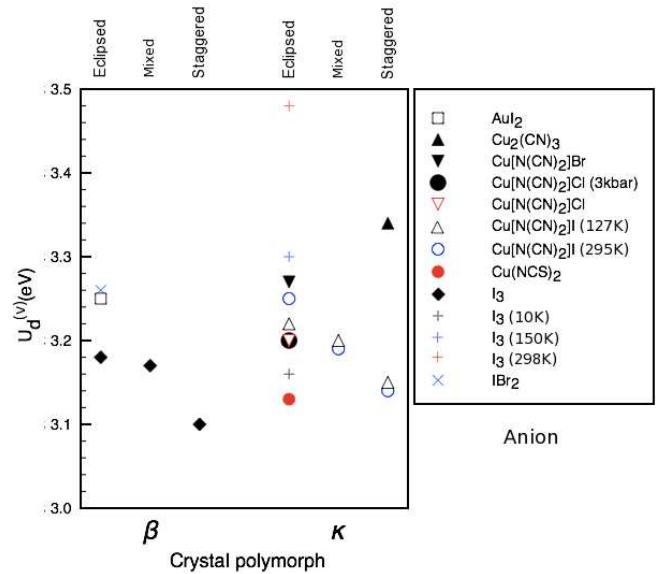, width=8.6cm}
\caption{The effective intra-dimer Coulomb repulsion, \udvn, for various ET dimers. The $x$-axis separates the data by source crystal polymorph ($\beta$ or $\kappa$), and by the terminal ethylene group conformation of each ET molecule in the dimer. \udv does not change significantly across the different ET crystals examined. For $\beta$-(ET)$_2$X crystals, \udv $ = 3.19 \pm 0.07$ eV. For $\kappa$-(ET)$_2$X crystals, \udv $= 3.23 \pm 0.09$ eV. The difference in \udv between the two crystal polymorphs is $\sim1 \%$. Therefore, there is no significant dependence of \udv on the dimer geometry associated with different crystal polymorphs.}\label{fig:Ud}
\end{figure} 

\subsection{Variation of the intra-dimer Coulomb repulsion}

\begin{figure}
\epsfig{figure=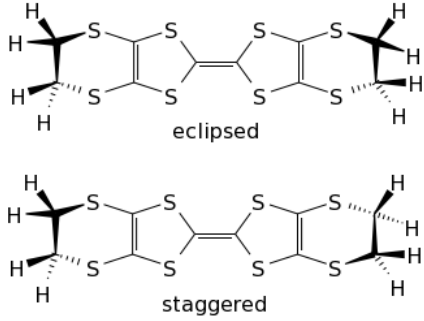, width=6cm}
\caption{ET molecules within the crystals studied occur in two conformations, denoted eclipsed and staggered. The difference between them lies in the relative orientation of the terminal ethylene groups.}\label{fig:2d_conformations}
\end{figure}

Now we consider variation of the Hubbard model parameters across the conformations found in different materials, beginning with \udvn. In Fig. \ref{fig:Ud} we show the values of \udv for the conformations observed experimentally in a variety of ET crystals. Of particular note are the three data points corresponding to different possible conformations of $\beta$-(ET)$_2$I$_3$. In the ET molecule the terminal ethylene groups may take two relative orientations known as the staggered and eclipsed conformations (cf. Fig. \ref{fig:2d_conformations}). \udv is smallest when both ET molecules are in the staggered conformation. Conversely, the largest \udv for this crystal occurs when both ET molecules are eclipsed, with intermediate \udv values for the case with one staggered and one eclipsed ET molecule. This trend is repeated in the $\kappa$-phase crystals, where two data sets (corresponding to different temperatures at which the nuclear positions were determined) for $\kappa$-(ET)$_2$Cu[N(CN)$_2$]I provide data for both conformations.

The mean value of \udv for the $\beta$ phase crystals is 3.19 $\pm$ 0.07 eV, and the mean \udv for the $\kappa$ phase crystals is 3.23 $\pm$ 0.09 eV. The quoted error is one standard deviation over the full set of conformations studied. The difference between the two values of \udv is $\sim 1\%$, and well within the error ranges. This suggests that \udv takes the same value, 3.22 $\pm$ 0.09 eV, in all $\beta$ and $\kappa$ phase ET salts. This result is significantly larger than the value of \udv obtained from H\"{u}ckel calculations ($\sim$0.5--2 eV), as we will discuss below.

\begin{figure}
\epsfig{figure=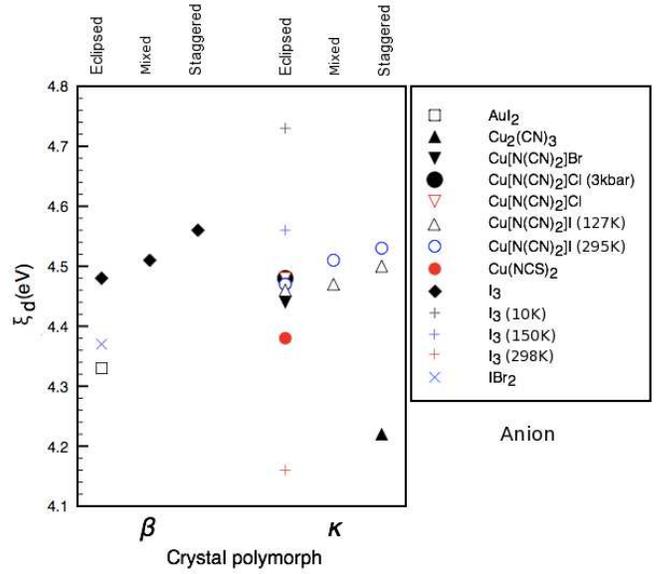, width=8.6cm}
\caption{Dimer hole site energy, \xidn, for various ET dimers. For $\beta$-(ET)$_2$X crystals, the mean value is \xid $= 4.45 \pm 0.10$ eV. For $\kappa$-(ET)$_2$X crystals, the mean value is \xid $= 4.46 \pm 0.14$ eV. The mean value for the whole data set is \xid $= 4.45 \pm 0.13$ eV. The effect of conformation on \xid is significantly larger for $\beta$-(ET)$_2$I$_3$ ($\sim 10\%$) than it is for the other parameters. The variations of \xid with dimer geometry associated with crystal polymorph and anion are $\sim 3\%$, similar to the relative variations of \udv and $V_m^{(v)}$ across the whole data set.}\label{fig:Xi}
\end{figure}

\subsection{Variations in site energy and the role of disorder}

As reviewed in the introduction, a number of experiments have shown that disorder has strong effects on both the normal state and superconducting properties of organic charge transfer salts.\cite{Powell:disorder,Analytis,Taylor,Taniguchi,Stalcup,Su,Su2,Wolter} There has been relatively little work on the effect of the random $U$  Hubbard model. Conclusions drawn from studies in one dimension\cite{Ugajin, Sandvik} cannot be straightforwardly generalised to higher dimensions. Mutou\cite{Mutou} used dynamical mean field theory to study the metallic phase of the random $U$ Hubbard model. However, he did not consider the effect a random $U$ on either superconductivity or the Mott transition, which are the primary concerns in the organic charge transfer salts. However, Mutou concluded that for small impurity concentrations Kondo-like effects  mean that the random $U$ Hubbard model is significantly different from the virtual crystal approximation to the random $U$ Hubbard model, which describes the system in terms of an average $U$. The only study\cite{Litak} we are aware of that discusses superconductivity  in the random  $U$ Hubbard model treats the negative $U$ model, which is not realistic for the organic charge transfer salts. Litak and Gy\"orffy\cite{Litak} studied a model where some sites have $U=0$ and others have a negative $U$. They find that superconductivity is suppressed above at certain critical concentration of $U=0$ sites. Therefore, it is not clear what implications our finding of small changes in \udv and hence \ud has for the physics of the organic charge transfer salts. However, it is interesting to ask what role this plays in the observed role of disorder in suppressing superconductivity\cite{Powell:disorder} and driving the system towards the Mott transition.\cite{Taniguchi}

\begin{figure}
\epsfig{figure=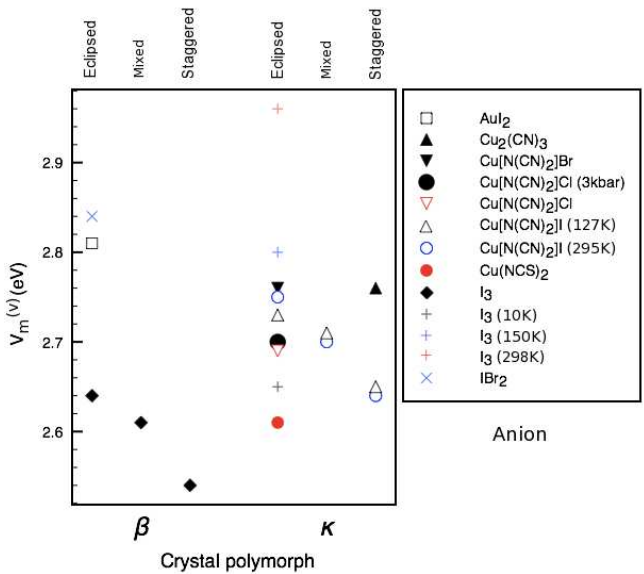, width=8.6cm}
\caption{Intradimer $V_m^{(v)}$ for various ET dimers.  For $\beta$-(ET)$_2$X crystals, $V_m^{(v)} = 2.69 \pm 0.13$ eV and for $\kappa$-(ET)$_2$X crystals, $V_m^{(v)} = 2.72 \pm 0.09$ eV. The mean value is $V_m^{(v)} = 2.71 \pm 0.10$ eV. The difference in $V_m^{(v)}$ between the crystal polymorphs is $\sim 2\%$. Therefore, $V_m^{(v)}$, like \udvn, does not significantly depend on the geometry associated with crystal polymorph. The effect of ET conformation on the value of $V_m^{(v)}$ in the crystals $\beta$-(ET)$_2$I$_3$ and $\kappa$-(ET)$_2$Cu[N(CN)$_2$]I is also similar to the effect on \udvn. $V_m^{(v)}$ is lowest when the ET dimer has the staggered-staggered conformation, and rises when either or both ET molecules are eclipsed. }\label{fig:V}
\end{figure} 

To understand the role of conformational disorder in terms of an effective Hamiltonian built up from ET dimers one must also understand the effect of conformational disorder of the effective dimer site energy (for holes), $\xi_d$. This is straightforwardly found from the DFT calculations described above via Eq. (\ref{eqn:xid}) and the results are reported in Fig. \ref{fig:Xi}. The effective scattering rate due to conformational disorder is given by
\begin{eqnarray}
\frac{\hbar}{\tau}=\sum_iN_i\pi D(E_F) |\Delta_i\xi_d|^2,
\end{eqnarray}
where $i$ labels the type of impurity (both staggered or mixed; the ground state conformation is both eclipsed), $N_i$ is number of impurities of type $i$, $D(E_F)$ is the density of states at the Fermi level, and $\Delta_i\xi_d$ is the difference between $\xi_d$ for $i$ type impurities and $\xi_d$ of eclipsed dimers.
 
In quasi-2D systems, $D(E_F)$ is simply related to the cyclotron electron mass\cite{Merino} by the relation
\begin{equation}
D(E_F) = \frac{m_c}{2 \pi \hbar^2}
\end{equation}
and in the presence of interactions Luttinger's theorem\cite{Luttinger} for a Fermi liquid produces
\begin{equation}
D(E_F) = \frac{m^*}{2 \pi \hbar^2}
\end{equation}
where $m^*$ is the effective mass. From quantum oscillation measurements, Wosnitza \etaln\cite{Wosnitza} found that $m^*/m_e$ = 4.2 in \bIn, where $m_e$ is the electron rest mass. From Shubnikov-de Haas measurements in \Br Caulfield \etal\cite{Caulfield} found that $m^*/m_e$ = 6.4. The scattering rate $\tau$ can be found from measurement of the interlayer residual resistivity, $\rho_0$, by the relation\cite{McKenzie:rho-tau}
\begin{equation}
\rho_0 = \frac{\pi \hbar^4}{2e^2m^*ct_\perp^2\tau}
\end{equation}
where $c$ is the interlayer lattice constant taken from the relevant x-ray scattering measurements\cite{kNCN2X:Geiser,bI3:Leung} and $t_\perp$ is the interlayer hopping integral, which has previously been estimated from experimental data for both \Br (Ref. \onlinecite{Wosnitza}) and \bI (Ref. \onlinecite{Powell:disorder}). Using these parameters we calculated the scattering rate in both the $\beta_L$ and $\beta_H$ phases of \bI from the low temperature values of $\rho_0$ reported by Ginodman \etaln.\cite{Ginodman} The scattering rate due to conformational impurities, $\tau_c^{-1}$ is then $\tau_c^{-1}=\tau_H^{-1}-\tau_L^{-1}$, where $\tau_L$ ($\tau_H$) is the quasiparticle lifetime in the $\beta_L$ ($\beta_H$) phase. Given our  calculated values of $\Delta_i\xi_d$  an $\sim 8\%$ concentration of staggered impurities would be required to cause this scattering rate. From a similar calculation comparing the residual resistivity measured in a single sample of \Br cooled at different rates we find that a $\sim 2\%$ concentration of staggered impurities would be sufficient to explain the rise increase in residual resistivity observed in the experiment utilising the fastest cooling over that performed with the slowest cooling rate. 
 X-ray scattering experiments\cite{Wolter} find that $3\pm3\%$ of the ET molecules are in the staggered conformation at 9 K, which is entirely consistent with our result. However, Wolter \etaln's\cite{Wolter} argument that this impurity concentration is too small to cause the observed effects of disorder in not sustained by the above calculations. Rather we find that all of the suppression in $T_c$ is entirely consistent with this degree of disorder.

\subsection{Variations in inter-molecular Coulomb repulsion}

In Fig. \ref{fig:V} we show the distribution of the calculated values of $V_m^{(v)}$.  The mean value of $V_m^{(v)}$ for the $\beta$ phase crystals is 2.69 $\pm$ 0.13 eV, while the mean value for the $\kappa$ phase crystals is 2.72 $\pm$ 0.09 eV. Again, the difference between the values is small ($\sim 2\%$) compared to the distribution for each polymorph. Therefore, $V_m^{(v)}$ is essentially the same across all of the conformations studied, with a mean value of 2.71 $\pm$ 0.10 eV.

\begin{table}
\begin{tabular}{lccc}
\hline
Crystal & Method & \udv (eV) \\
\hline
$\beta$-(ET)$_2$I$_3$ & H\"uckel\cite{H-lit:Mori} & 0.49\\
$\beta$-(ET)$_2$IBr$_2$ & H\"uckel\cite{Emge} & 0.98\\
$\beta$-(ET)$_2$ICl$_2$ & H\"uckel\cite{Emge} & 1.04\\
$\beta$-(ET)$_2$I$_3$ & H\"uckel\cite{H-lit:Mori} & 0.49\\
$\beta$-(ET)$_2$CH(SO$_2$CF$_3$)$_2$ & H\"uckel\cite{H-lit:Schlueter} & 0.88-0.90 \\
$\beta$-(ET)$_2$[OsNOCl$_5$] & H\"uckel\cite{H-lit:Simonov} & 2.10 \\
$\kappa$-(ET)$_2$Cu[N(CN)$_2$]Cl & DFT\cite{Kandpal:DFT_band} & 0.4 \\
$\kappa$-(ET)$_2$Cu[N(CN)$_2$]Br & H\"uckel\cite{H-lit:Fortunelli} & 0.45 \\
$\kappa$-(ET)$_2$Cu(NCS)$_2$ & H\"uckel\cite{H-lit:Campos} & 0.48 \\
$\kappa$-(ET)$_2$Cu(NCS)$_2$ & H\"uckel\cite{kNCS2:Rahal} & 0.14 \\
$\kappa$-(ET)$_2$Cu(NCS)$_2$ & H\"uckel\cite{H-lit:Komatsu} & 0.46 \\
$\kappa$-(ET)$_2$Cu(NCS)$_2$ & DFT\cite{Nakamura:DFT_band} & 0.83 \\
$\kappa$-(ET)$_2$Cu(NCS)$_2$ & DFT\cite{Kandpal:DFT_band} & 0.4 \\
$\kappa$-(ET)$_2$Cu[N(CN)$_2$]Br & H\"uckel\cite{H-lit:Komatsu} & 0.49 \\
$\kappa$-(ET)$_2$Cu$_2$(CN)$_3$ & H\"uckel\cite{H-lit:Komatsu} & 0.45 \\
$\kappa$-(ET)$_2$Cu$_2$(CN)$_3$ & DFT\cite{Nakamura:DFT_band} & 0.85 \\
$\kappa$-(ET)$_2$Cu$_2$(CN)$_3$ & DFT\cite{Kandpal:DFT_band} & 0.4 \\
$\kappa$-(ET)$_2$I$_3$ & H\"uckel\cite{H-lit:Komatsu} & 0.49 \\
$\kappa$-(ET)$_2$I$_3$ & H\"uckel\cite{kI3:Kobayashi} & 0.22 \\
\hline
\end{tabular}
\caption{Previous estimates of \udv for various $\beta$- and $\kappa$-phase ET salts. These values were obtained from both H\"uckel and density functional methods under the assumptions $U_m^{(v)} \rightarrow \infty$ and $V_m^{(v)} = 0$, which yields $U_d^{(v)}=2t$. These estimates substantially underestimate the actual value of \udv (see Fig. \ref{fig:Ud}) as $U_m^{(v)}\sim V_m^{(v)}$. The two site extended Hubbard model produces values of $t$ on the same order of magnitude as these H\"{u}ckel calculations. One should also note the wide scatter between the different H\"uckel calculations, even between different studies of the same material. }\label{tab:t_lit}
\end{table}

Previous calculations of \udv based on both the Huckel method\cite{H-lit:Fortunelli,Emge,H-lit:Mori,H-lit:Schlueter,H-lit:Simonov,H-lit:Campos,kNCS2:Rahal,H-lit:Komatsu,kI3:Kobayashi} and DFT\cite{Kandpal:DFT_band,Nakamura:DFT_band} have assumed that  $U_m^{(v)} \rightarrow \infty$ and $V_m^{(v)} = 0$. Substituting these conditions into Eqs. (\ref{eqn:Ud}) and (\ref{eqn:simpleEvals}) yields \udv $= 2t$. Literature values of \udv based on this approximation are presented in Table \ref{tab:t_lit} for comparison with our DFT results. It can be seen that this assumption yields values of \udv that are significantly smaller than those we have calculated above (cf. Fig. \ref{fig:Ud}). 
However, we have previously found\cite{Scriven:monomers} that $U_m^{(v)}=4.2 \pm0.1$ eV.
Comparing this with the above  results we see that ${U_m^{(v)}}/{V_m^{(v)}} \sim 1.5$, in contradiction to the assumption that $U_m^{(v)} \gg V_m^{(v)}$. Hence $U_m^{(v)}\gg|2t|$.

If we instead make the assumption $U_m^{(v)} \simeq V_m^{(v)} \gg |t|$, then Eqs. (\ref{eqn:Ud}) and (\ref{eqn:simpleEvals}) give
\begin{equation}
U_d^{(v)} \approx \frac{1}{2}(U_m^{(v)} + V_m^{(v)}).
\end{equation}
Substituting in the mean values of $U_m^{(v)}$ and $V_m^{(v)}$ gives \udv = 3.41 eV. This result is close (within $6\%$) to our calculated value of \udvn. Therefore, this is a reasonable approximation for the ET salts. Further, this shows that the result that \udv does not vary significantly because of changes in conformation between different salts or polymorphs is a consequence of the fact that neither \umv or $V_m^{(v)}$ vary significantly because of changes in conformation between different salts or polymorphs.

\section{Conclusions}
The effective Coulomb repulsion terms in the Hubbard model are essentially the same for all of the ET conformations studied. We found that \udv $= 3.22 \pm 0.09$ eV, $V_m^{(v)} = 2.71 \pm 0.10$ eV. The value of \udv is significantly larger than previous estimates based on the extended H\"{u}ckel formalism or DFT under the assumptions $U_m^{(v)} \rightarrow \infty$ and $V_m^{(v)} = 0$. This can be understood because we have shown that $U_m^{(v)} \sim V_m^{(v)}$ and hence $U_d^{(v)} \approx \frac{1}{2}(U_m^{(v)} + V_m^{(v)})$.

The lack of variation of \udv between the two polymorphs and when the anion is changed is interesting in the context of theories of these organic charge transfer salts based on the Hubbard model. These theories require $U_d/W$ to vary significantly as the anion is changed (chemical pressure) and under hydrostatic pressure. Therefore our results show that either  $U_d^{(p)}$ or $W$ must vary significantly under chemical and hydrostatic pressure, or else these theories do not provide a correct description of the $\beta$ and $\kappa$ phase organic charge transfer salts. This is particularly interesting as fast cooling has been shown to drive \Br to the insulating side of the metal-insulator transition.\cite{Taniguchi} 

We have also studied the effects of conformational disorder on these parameters, which is found to be quite small, consistent with the often subtle effects of conformational disorder observed in these materials. The largest changes are found in the geometries taken from $\beta$-(ET)$_2$I$_3$, which shows the strongest effects of conformational disorder. It is also interesting that we found a systematic variation in \udv is caused by conformational disorder. As there has been relatively little work on the random $U$ Hubbard model it is difficult to speculate what effects this has on the low temperature physics of the organic charge transfer salts at present.

Given that DFT band structure parameterisations of the interdimer hopping integrals have recently been reported for several organic charge transfer salts,\cite{Kandpal:DFT_band,Nakamura:DFT_band} the outstanding challenge for the parameterisation of the Hubbard model in these systems is the accurate calculation of $U_d^{(p)}$. The bandwidth in both the $\beta$ (Ref. \onlinecite{Lee:DFT_band}) and $\kappa$ (Ref. \onlinecite{Kandpal:DFT_band,Nakamura:DFT_band}) phase salts is around 0.4-0.6 eV. Therefore, our finding that \udv is significantly larger than has been realised previously shows that $U_d^{(p)}$ must be significant as if $U_d\simeq U_d^{(v)}$ then all of these materials would be well into the Mott insulating regime.  Thus $U_d^{(p)}$ must significantly reduce $U_d$ in order for the, observed, rich phase diagram to be realised. This is consistent with comparisons of DMFT calculations to optical conductivity measurements on $\kappa$-(ET)$_2$Cu[N(CN)$_2$]Br$_x$Cl$_{1-x}$, which suggest that $U_d = 0.3$ eV.\cite{Merino:Ud} Further, $U_d^{(p)}$ may be quite sensitive to the crystal lattice and therefore may be important for understanding the strong dependence of these materials on chemical and hydrostatic pressure.

\acknowledgements

We thank Ross McKenzie for helpful comments on a draft of this manuscript. This work was supported by the Australian Research Council (ARC) under the Discovery scheme (Project No. DP0878523) and by a University of Queensland Early Career Research grant. B.J.P. was supported by the ARC under the Queen Elizabeth II scheme. Numerical calculations were performed on the APAC national facility under a grant from the merit allocation scheme.

\end{document}